\documentclass[runningheads]{llncs}
\usepackage[T1]{fontenc}
\usepackage{graphicx}
\usepackage{bm}
\usepackage{amsmath}
\usepackage{amssymb}
\usepackage{amsfonts}
\usepackage{hyperref}
\usepackage{graphics}
\usepackage{tabularx}
\usepackage{booktabs}
\usepackage{color}
\usepackage{multirow}
\usepackage{siunitx}
\usepackage{appendix}
\usepackage[normalem]{ulem}
\usepackage[misc,geometry]{ifsym}
\begin{document}

\title{Deep-learning-based groupwise registration for motion correction of cardiac $T_1$ mapping}
\titlerunning{Groupwise registration for cardiac T1 mapping}

\author{Yi Zhang \inst{1}
\and Yidong Zhao\inst{1}
\and Lu Huang\inst{2}
\and Liming Xia\inst{2}
\and Qian Tao\inst{1} (\Letter)
} 
%index{Zhang,Yi} 
%index{Zhao,Yidong}
%index{Huang,Lu}
%index{Xia,Liming}
%index{Tao,Qian}

\authorrunning{Y. Zhang et al.}
\institute{Department of Imaging Physics, Delft University of Technology, The Netherlands
\email{q.tao@tudelft.nl}
\and Tongji Medical College, Huazhong University of Science and Technology, China}

\maketitle              
\begin{abstract}
Quantitative $T_1$ mapping by MRI is an increasingly important tool for clinical assessment of cardiovascular diseases. The cardiac $T_1$ map is derived by fitting a known signal model to a series of baseline images, while the quality of this map can be deteriorated by involuntary respiratory and cardiac motion. To correct motion, a template image is often needed to register all baseline images, but the choice of template is nontrivial, leading to inconsistent performance sensitive to image contrast. In this work, we propose a novel deep-learning-based groupwise registration framework, which omits the need for a template, and registers all baseline images simultaneously. We design two groupwise losses for this registration framework: the first is a linear principal component analysis (PCA) loss that enforces alignment of baseline images irrespective of the intensity variation, and the second is an auxiliary relaxometry loss that enforces adherence of intensity profile to the signal model. We extensively evaluated our method, termed ``PCA-Relax'', and other baseline methods on an in-house cardiac MRI dataset including both pre- and post-contrast $T_1$ sequences. All methods were evaluated under three distinct training-and-evaluation strategies, namely, standard, one-shot, and test-time-adaptation. The proposed PCA-Relax showed further improved performance of registration and mapping over well-established baselines. The proposed groupwise framework is generic and can be adapted to applications involving multiple images.

\end{abstract}
%
%
%
%%%%%%%%%%%%%%%%%%%%%%%%%%%%%%%%%%%%%%%%%%%%%%%%%%
%%%%%%%%%%%%%%%%%%% Introduction %%%%%%%%%%%%%%%%%
%%%%%%%%%%%%%%%%%%%%%%%%%%%%%%%%%%%%%%%%%%%%%%%%%%
\section{Introduction}

Quantitative MRI (qMRI) of heart, such as $T_1$ and $T_2$ mapping~\cite{messroghli2004modified,o2022t2}, has become an increasingly important imaging modality for non-invasive examination of heart diseases~\cite{haaf2016cardiac}. In principle, the quantitative values of $T_1$ or $T_2$ relaxometry are inferred by fitting a known parametric model to a series of baseline images with different intensities and contrasts, assuming anatomical alignment. In practice, however, the alignment assumption is often violated by the involuntary respiratory and cardiac motion of patients~\cite{xue2012motion}, resulting in deteriorated accuracy and precision of the quantitative mapping~\cite{kellman2013t1,kellman2014t1}. This makes motion correction an essential post-processing step for qMRI~\cite{makela2002review,tao2018robust}. 

Conventionally, qMRI motion correction is performed in a pairwise fashion, by first selecting a single image or an average image as the \textit{template}, and then registering the rest of the series to this template~\cite{bron2013image}. The choice of template, however, is nontrivial~\cite{li2021learning,qiao2019fully}. Some baseline images of qMRI can have extremely poor contrast depending on the acquisition setting and can fail catastrophically in pairwise registration. Such failure can severely undermine the final mapping quality. In contrast, groupwise registration, which registers all baseline images simultaneously, is a promising alternative. The groupwise paradigm avoids explicit selection of a template, and utilizes the shared information among all baseline images. The design of groupwise similarity, however, is not as straightforward as pairwise similarity. Aggregation of pairwise metrics was proposed to describe the groupwise alignment, including accumulated pairwise estimates~\cite{wachinger2012simultaneous} and the sum of variances~\cite{metz2011nonrigid}. Notably, Huizinga \textit{et al.}~\cite{huizinga2016pca} proposed a principle component analysis (PCA)-based metric that characterizes groupwise alignment without aggregating pairwise metric computation, in the Elastix framework~\cite{klein2009elastix}.

The role of image registration in medical image analysis is well-established, with traditional optimization-based methods that optimize the deformation fields per dataset~\cite{bron2013image,huizinga2016pca,klein2009elastix,makela2002review}, and recent deep-learning-based methods that predict the deformation fields through a parameterized network\cite{arava2021deep,voxelmorph,fu2020deep,hanania2023pcmc,li2021learning,li2023contrast,li2022motion,martin2020groupwise,yang2022end}. Compared with iterative optimization-based methods, deep-learning-based methods promises much faster inference with competitive performances~\cite{voxelmorph,fu2020deep}. Interestingly, deep-learning-based methods can also be interpreted as amortized optimization on the training dataset, generalizable to in-domain data~\cite{voxelmorph}. For qMRI, most deep-learning-based methods follow the pairwise paradigm, relying on the selection of a template~\cite{arava2021deep,li2022motion}. A groupwise deep learning framework for qMRI was proposed recently,~\cite{li2023contrast,zhang2021groupregnet}, but it still relies on the selection of a template and aggregates pairwise metrics as the groupwise loss.

With the MRI signal model known, physics-informed qMRI registration has been explored~\cite{tilborghs2019robust,van2013model,xue2012motion}. The methods are mostly optimization-based, which is typically slow, and the additional mapping loop further adds to the long computation. A recent work, PCMC-T1~\cite{hanania2023pcmc}, predicts motion-corrected baseline series by minimizing qMRI fitting error. However, the registration solely relied on the $T_1$ fitting error, which can be sensitive to the initial motion~\cite{tao2018robust}, while susceptible to shape collapse (\textit{i.e.}, overfitting the signal model) \cite{li2023contrast}.

In this work, to realize fast, template-free, physics-informed motion correction of $T_1$ mapping, we propose a novel groupwise registration framework. The groupwise registration makes use of PCA with a robust yet straightforward premise: the intensity profiles of all pixels should adhere to a low-rank model. This \textit{implicitly} regulates anatomical alignment across baseline images. Furthermore, we design an auxiliary relaxometry loss, which \textit{explicitly} incorporates the MR relaxometry into the registration, also in a groupwise manner. The second loss serves to refine the registration after PCA, given that it can be sensitive to motion and prone to overfit. We show that the proposed method, termed ``PCA-Relax'', significantly outperformed established medical image registration baselines in our extensive experiments with different training-and-evaluation settings.

%%%%%%%%%%%%%%%%%%%%%%%%%%%%%%%%%%%%%%%%%%%%%%%%%%
%%%%%%%%%%%%%%%%%%%%%% Method %%%%%%%%%%%%%%%%%%%%
%%%%%%%%%%%%%%%%%%%%%%%%%%%%%%%%%%%%%%%%%%%%%%%%%%

\section{Method}

\subsection{Groupwise Image Registration}
\begin{figure}[htbp]
    \centering
    \includegraphics[scale = 0.9]{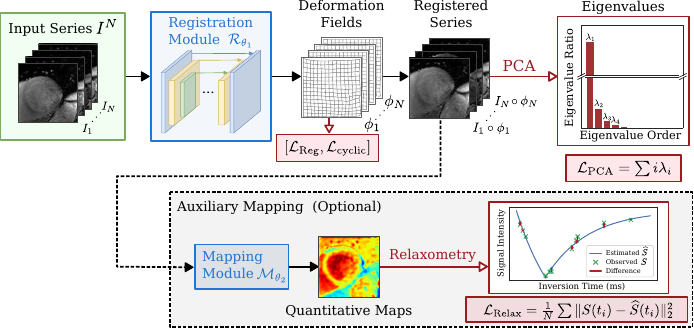}
    \caption{An overview of the proposed template-free groupwise registration framework. It takes the baseline image series $I^N$ through the registration module to predict $\phi^N$. The warped series $I^N \circ \phi^N$ undergoes PCA decomposition to calculate the groupwise PCA loss without a template. If the auxiliary mapping module is enabled, the relaxometry loss is also activated to refine the registration module.}
    \label{fig:pipeline}
\end{figure}

In quantitative MRI, a series of $N$ baseline images are acquired for mapping, denoted by $I^N = \left\{ I_i  | i = 1, 2, \ldots, N \right\}$, where $ I_i \in \mathbb{R}^{H \times W}$ is an image of size $H \times W$. The objective of groupwise registration is to align these images into a common coordinate system through parameterized 2D deformation fields $\phi^N = \{\phi_i | i =1,2,\ldots, N \}$, ensuring that all $I^N \circ \phi^N = \{I_i\circ \phi_i | i = 1,2,\ldots, N \}$ align. The proposed deep-learning groupwise image registration framework is built upon the well-established VoxelMorph backbone~\cite{voxelmorph}. As shown in Fig. \ref{fig:pipeline}, the parameterized network $\mathcal{R}_{\theta_1}$ takes $I^N$ stacked along the channel dimension as input, and passes them through a U-Net architecture~\cite{ronneberger2015u} to predict $\phi^N$. 

\subsection{PCA-based Template-free Similarity Metric}
\begin{figure}[htbp]
    \centering
    \includegraphics[scale = .85]{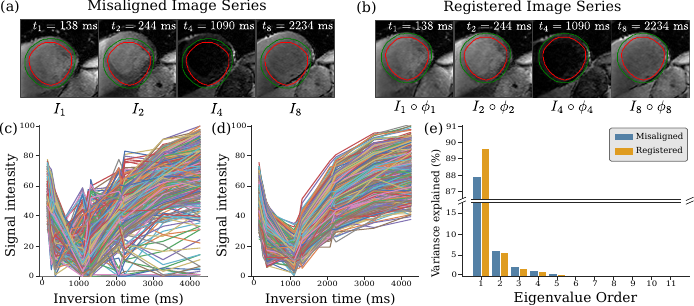}
    \caption{Illustration of cardiac motion correction in qMRI. Sampled (a) misaligned and (b) registered series. Voxel-wise intensity curve for (c) misaligned and (d) registered series. And (e) the comparison of eigenvalues of the correlation matrix.}
    \label{fig:teaser}
\end{figure}

In qMRI, the signal intensity $S_{(x,y)}$ at image coordinate $(x,y)$, follows a parametric signal model. For $T_1$ mapping, with the widely adopted Modified Look-Locker inversion recovery (MOLLI)~\cite{messroghli2004modified} sequence, the signal model is defined as follows:

\begin{align}
    S_{(x,y)}(t_i) &= \left \vert C_{(x,y)} \left(1 -  k_{(x,y)} \exp\left(-\frac{t_i}{{T_1^*}_{(x,y)}}\right)\right) \right \vert,
    \label{eq:molli}
\end{align}
where $t_i$ is the inversion time of $I_i$. The parameters $C_{(x,y)}$, $k_{(x,y)}$ and ${T_1^*}_{(x,y)}$ are the underlying tissue parameters at coordinates $(x,y)$ to derive ${T_1}_{(x,y)} = (k_{(x,y)}-1){T_1^*}_{(x,y)}$. For a well-aligned MOLLI image series, the group of signal profiles has a low rank. This can be visually appreciated, as shown in Fig. \ref{fig:teaser} (d).

PCA provides an intuitive way to evaluate the groupwise alignment across all baseline images. By first rearranging the registered series $I^N \circ \phi^N$ into a data matrix $M \in \mathbb{R}^{HW \times N}$ such that each row represents a signal profile of length $N$, the alignment of all baseline images can be characterized by performing PCA decomposition on the normalized correlation matrix $K$: 

\begin{align}
    K&=\frac{1}{HW-1} \Sigma^{-1}(M-\overline{M})^{\top}(M-\overline{M}) \Sigma^{-1},\label{eq:correlation}\\
    K& =U\Lambda U^{\top}, \quad  \Lambda = \mathrm{diag}(\lambda_1, \lambda_2, \ldots, \lambda_N),
    \label{eq:pca}
\end{align}
where $\Sigma = \mathrm{diag}(\sigma_1,\sigma_2, \ldots,\sigma_N)$ with $\sigma_i$ being the standard deviation of each column, $\overline{M}$ is the column-wise mean matrix, $U$ is the orthogonal matrix of eigenvectors, and $\lambda_1, \lambda_2,\ldots, \lambda_N$ are the eigenvalues of $K$ in descending order. Ideally, the variance of $K$ should be explained by the first few eigenvalues if all baseline images are aligned. However, motion can induce noisy or shifted row entries in $M$, resulting in a more scattered distribution of eigenvalues of $K$  (Fig. \ref{fig:teaser} e). Therefore, we define the PCA loss as follows~\cite{huizinga2016pca}:
\begin{equation}
    \mathcal{L}_{\text{PCA}} = \sum_{i=1}^N i\lambda_i.
\end{equation}
Since $\sum_{i=1}^N \lambda_i$ equals to $N$ given the normalized correlation matrix $K$, a smaller $\mathcal{L}_{\text{PCA}}$ indicates a sharper energy concentration at the first few eigenvalues. 

\subsection{Auxiliary Relaxometry Loss}
In principle, the aligned series should follow the MR signal model at each voxel, \textit{e.g.}, as in Eq. \ref{eq:molli}. However, quantitative parameters $\left(C, k, T_1^* \right)$ are normally estimated by least-square methods or grid search~\cite{van2013model,tilborghs2019robust}, lacking practical differentiability. This makes it difficult to integrate the physics prior into the deep learning registration framework. To incorporate the physics prior of relaxometry, we propose a \textit{differentiable} qMRI mapping module and use it to further refine the PCA-based registration. We design an end-to-end U-Net~\cite{ronneberger2015u} architecture to construct this module $\mathcal{M}_{\theta_2}$, parameterized by ${\theta_2}$~\cite{zhao2023relaxometry}. The module takes stacked registered images $I^N \circ \phi^N$ and inversion times $t^N = \left\{t_i \in \mathbb{R}^{H \times W}| i = 1,2,\ldots, N \right\}$ as input, and output parameter maps $\left[C, k, T_1^* \right] \in \mathbb{R}^{3\times H \times W}$. The mapping module is pre-trained, independently of the registration module, in a fully self-supervised fashion. The relaxometry loss is defined as the normalized fitting error:
\begin{equation}
    \mathcal{L}_{\text{Relax}} = \frac{1}{N HW}\sum_{(x,y)\in \Omega}\sum_{i=1}^{N}\left\Vert S_{(x,y)}(t_i) - \widehat{S}_{(x,y)}(t_i)\right\Vert_2^2,
    \label{eq:relax_loss}
\end{equation}
where $\Omega$ denotes the spatial domain of an image $I \in \mathbb{R}^{H \times W}$, $S_{(x,y)}(t_i)$ is the signal intensity of $I_i \circ \phi_i$ at $(x,y)$, and $\widehat{S}_{(x,y)}(t_i)$ is the estimated intensity by evaluating Eq. \ref{eq:molli} at $(x,y)$. The differentiability of $\mathcal{L}_{\text{Relax}}$  is established through its sequential composition of $I^N \circ \phi^N$, $\mathcal{M}_{\theta_2}$, and signal model. We note that in our work, $\mathcal{L}_{\text{Relax}}$ is \textit{optional} as shown in Fig. \ref{fig:pipeline}, and can be omitted if the registration problem is model-agnostic. 

\subsection{Regularization on Groupwise Deformation} 

In addition to the two template-free groupwise losses $\mathcal{L}_\text{PCA}$ and $\mathcal{L}_{\text{Relax}}$, we further regularize the deformation fields as regularly done. The first is the regularization loss $\mathcal{L}_{\text{reg}}$~\cite{voxelmorph} on deformation fields to ensure the spatial smoothness of $\phi^N$:
\begin{equation}
    \mathcal{L}_{\text{reg}} = \frac{1}{NHW} \sum_{(x,y) \in \Omega} \sum_{i=1}^{N} \left\Vert\nabla \phi_i(x,y)\right\Vert_2^2,
    \label{eq:reg}
\end{equation}
where $\nabla$ denotes the spatial gradient operator. 

Specific to groupwise registration is the cyclic consistency loss $\mathcal{L}_{\text{cyclic}}$ to prevent collapsing~\cite{huizinga2016pca,li2021learning,zhang2021groupregnet}:
\begin{equation}
\mathcal{L}_{\text {cyclic }}=\sqrt{\frac{1}{NHW} \sum_{(x, y) \in \Omega}\left(\sum_{i=1}^{N} \phi_i(x, y)\right)^2},
\end{equation}
which enforces the deformation fields to warp the baseline images to a ``mean shape'' of the group~\cite{huizinga2016pca,qiao2019fully}. Therefore, the total loss is:
\begin{equation}
    \label{eq:total loss}
    \mathcal{L}_{\text{total}} = \lambda_{\text{PCA}}\mathcal{L}_{\text{PCA}} +  \lambda_{\text{reg}}\mathcal{L}_{\text{reg}} + \lambda_{\text{cyclic}}\mathcal{L}_{\text{cyclic}} + \lambda_{\text{Relax}}\mathcal{L}_{\text{Relax}}.
\end{equation}

%%%%%%%%%%%%%%%%%%%%%%%%%%%%%%%%%%%%%%%%%%%%%%%%%%
%%%%%%%%%%%%%%%%%%%% Experiments %%%%%%%%%%%%%%%%%
%%%%%%%%%%%%%%%%%%%%%%%%%%%%%%%%%%%%%%%%%%%%%%%%%%
\section{Experiments and Results}
\noindent\textbf{Data:} We used an in-house cardiac MRI dataset including 50 subjects. Each subject has both pre-contrast and post-contrast MOLLI sequences (Philips 3.0T) with a fixed length $N = 11$. %The values of inversion times $t^N$ vary across each series. 
All images were resampled to a $1$ mm resolution and center-cropped to a size of $128 \times 128$. The dataset was randomly split subject-wise to prevent data leakage: $(36,4,10)$ subjects were used for training, validation, and testing. To evaluate out-of-domain generalization, training involved only pre-contrast sequences, whereas validation and testing included both pre-contrast and post-contrast sequences.

\noindent\textbf{Training-and-evaluation:} We employed three train-and-evaluation settings to assess the registration performance. \textbf{Standard}: the model was trained on the training dataset and unchanged at inference time. \textbf{One-shot}: the model was randomly initialized and trained on the input $T_1$ mapping sequence. \textbf{Test-time Adaptation (TTA)}: the model was pre-trained on the training dataset and finetuned on the input image as in~\cite{voxelmorph}. 

\noindent\textbf{Comparison Study:}
The following scenarios were compared: \begin{enumerate}
    \item Raw: Original series without any registration.
    \item VM-P: A pairwise registration baseline with the VoxelMorph backbone, which registered all baseline images to $I_{\text{template}} = I_1$, with normalized mutual information (NMI) loss. 
    \item VM-G: A template-based groupwise registration baseline with the VoxelMorph backbone, using $I_{\text{template}} = \frac{1}{N}\sum_{i=1}^N I_i\circ \phi_i$ as the template, and aggregated pairwise NMI loss.
    \item PCA: Our template-free groupwise framework with $\mathcal{L}_{\text{PCA}}$.
    \item PCA-Relax: Our template-free groupwise framework with $\mathcal{L}_{\text{PCA}}$ and $\mathcal{L}_{\text{Relax}}$.
\end{enumerate}

\noindent{\textbf{Evaluation Metrics:}}
We evaluated our proposed method in terms of $T_1$ mapping quality, as indicated by the fitting SD values~\cite{kellman2013t1}. Instead of evaluating tissue heterogeneity, the fitting SD measures the quality of curve fitting of the signal model at each voxel. The SD map is a clinically accepted metric to evaluate the quality of $T_1$ mapping~\cite{haaf2016cardiac,kellman2014t1,schelbert2016state,tilborghs2019robust}, as it is difficult to compare contours or landmarks across the cardiac qMRI baselines given the varying contrast. We estimated the fitting SD following~\cite{kellman2013t1} and calculated the SD values in the myocardium region for each series, manually annotated by experienced radiologists.

\noindent\textbf{Implementation Details:}
Our models were developed in PyTorch, with thorough hyper-parameter tuning on the validation split. The hyperparameters for PCA-Relax were set to $\lambda_{\text{PCA}} = 1, \lambda_{\text{reg}} = 10, \lambda_{\text{cyclic}} = 0.1,\lambda_{\text{Relax}} = 10$. For PCA, we set $\lambda_{\text{Relax}} = 0$. The models VM-P and VM-G utilized the NMI loss with $\lambda_{\text{NMI}} = 10$. For mapping module $\mathcal{M}_{\theta_2}$, the encoder features 5 convolutional layers with channel counts $[32,32,32,64,64]$; the decoder mirrors the encoder. We pretrained $\mathcal{M}_{\theta_2}$ for 100 epochs on the training split. The architecture of $\mathcal{R}_{\theta_1}$ is the same as $\mathcal{M}_{\theta_2}$. We used the ADAM optimizer, with learning rates of $5 \times 10^{-4}$ for Standard, and $1 \times 10^{-3}$ for One-shot, and TTA with early stopping after 500 iterations. 
Parametric fitting was performed using the Nelder-Mead algorithm after registration. All experiments were conducted on an NVIDIA RTX 4090 GPU. Our code will be released on GitHub. 

%%%%%%%%%%%%%%%%%%%%%%%%%%%%%%%%%%%%%%%%%%%%%%%%%%
%%%%%%%%%%%%%%%%%%%% Conclusion %%%%%%%%%%%%%%%%%%
%%%%%%%%%%%%%%%%%%%%%%%%%%%%%%%%%%%%%%%%%%%%%%%%%%

\begin{figure}[h]
    \centering
    \includegraphics[scale = 1.0]{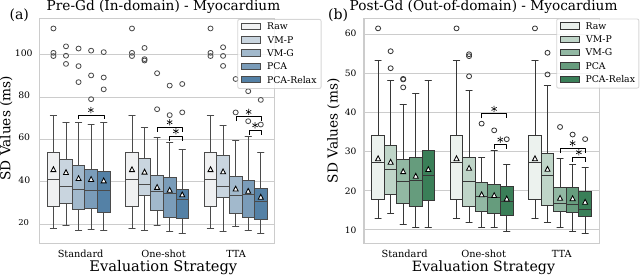}
    \caption{Boxplots of $T_1$ SD values in the myocardium, with lower values indicating better motion correction: (a) Pre-Gd (in-domain) and (b) Post-Gd (out-of-domain). All five scenarios were evaluated in three training-and-evaluation settings. One-sided Wilcoxon signed-rank tests were conducted to compare the performance of PCA-Relax against that of PCA and VM-G. Statistically significance ($p < 0.05$) is labeled with *.}
    \label{fig:box}
\end{figure}

\begin{figure}[htbp]
    \centering
    \includegraphics[scale= 1.0]{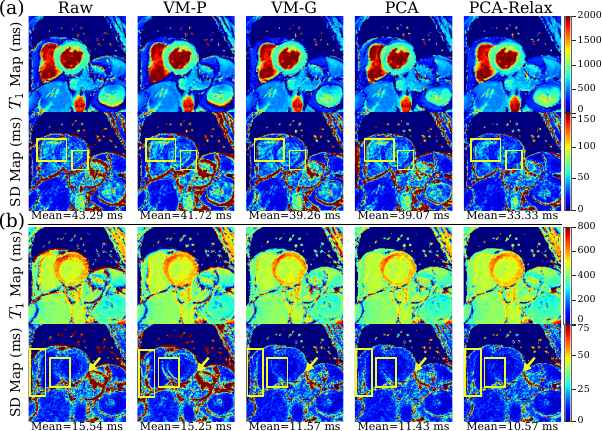}
    \caption{Estimated $T_1$ and SD maps of a (a) pre-contrast and (b) post-contrast sequence with the TTA strategy. The mean values of the SD maps in the myocardium are reported. We highlight the difference in the SD maps by the yellow boxes and arrows.}
    \label{fig:qualitative}
\end{figure}

\noindent\textbf{Results:} We evaluated all four motion correction methods, as well as the raw data, on three training-and-evaluation settings. The box plots are shown in Fig. \ref{fig:box}, with detailed statistics reported in Table \ref{tab:test_results}. Our template-free registration baseline (PCA) outperforms the competitive VoxelMorph baselines (VM-P, VM-G) in all scenarios. Furthermore, for all-but-one settings, the SD maps are improved with the mapping module activated (PCA-Relax). However, the post-Gd results in the standard setting suggest that the mapping module may overfit when a domain shift exists. (Note that the training only included the pre-contrast $T_1$ mapping data.) One-shot optimization and TTA per sequence take only $\approx 20$ secs with the groupwise framework, making TTA a valuable trade-off for refined registration for each new input sequence, with little extra time. Qualitative improvements in cardiac $T_1$ mapping are illustrated in Fig. \ref{fig:qualitative}. 

\begin{table}[htbp]

    \caption{Mean and standard deviation of the $T_1$ SD values in the myocardium, with three training-and-evaluation strategies. \textbf{Bold} denotes the best results within the specific training-and-evaluation strategy and \underline{underline} denotes the overall best results.}
    \centering    \resizebox{\textwidth}{!}{
    \begin{tabular}{p{2cm} p{1.8cm} m{2.2cm}<{\centering} m{2.3cm}<{\centering} m{2.3cm}<{\centering} m{2.3cm}<{\centering} m{2.3cm}<{\centering}}
        \toprule 
        \multicolumn{1}{l}{\multirow{2}{*}{Method}} & \multicolumn{1}{l}{\multirow{2}{*}{Modality}} &  \multicolumn{3}{c}{SD Values $(\si{ms})$ $\downarrow$} \\ 
         & & Standard &  One-shot & Fine-tuning\\
        \midrule
        Raw & Pre-Gd & 45.89 ($\pm$23.33) &  & \\
        VM-P &Pre-Gd& 44.50 ($\pm$21.56)&44.71 ($\pm$20.90)&44.84 ($\pm$21.33)\\
        VM-G  & Pre-Gd & 41.67 ($\pm$21.24) & 37.50 ($\pm$16.22) & 36.72 ($\pm$15.73) \\

        PCA  & Pre-Gd&  41.25 ($\pm$20.35) & 36.15 ($\pm$15.44) & 35.66 ($\pm$15.04) \\
        PCA-Relax & Pre-Gd & \textbf{40.58} ($\pm$20.95) & \textbf{34.00} ($\pm$15.89) & \textbf{\underline{32.88}} ($\pm$14.41)\\
    \midrule  
        Raw & Post-Gd & 28.32 ($\pm$12.15) & & \\
        VM-P &Post-Gd&27.42 ($\pm$10.79)&25.86 ($\pm$12.10)& 25.60 ($\pm$11.31)\\
        
        VM-G & Post-Gd&  25.00 ($\pm$10.28) & 19.09 ($\pm$5.75) & 18.19 ($\pm$5.46) \\
        PCA  & Post-Gd&  \textbf{23.82} ($\pm$\textbf{9.44}) & 18.89 ($\pm$5.85) & 18.08 ($\pm$5.28) \\
        PCA-Relax & Post-Gd & 25.55 ($\pm$10.04) &  \textbf{18.02} ($\pm$5.43) & \textbf{\underline{17.19}} ($\pm$5.28) \\
        \bottomrule

        \end{tabular}
        }
    \label{tab:test_results}
\end{table}

\section{Conclusion}
In this work, we proposed a novel template-free, deep-learning-based, groupwise registration framework, to tackle the motion correction problem for cardiac $T_1$ mapping. Two groupwise losses were proposed and validated: a sequence-agnostic PCA loss and a sequence-specific relaxometry loss. We extensively evaluated the proposed method, PCA-Relax, with diverse training-and-evaluation strategies on an in-house cardiac $T_1$ mapping dataset. The proposed method demonstrated improved performance of registration and mapping over well-established baselines. The generic formulation of our groupwise framework allows easy extension to applications that involve multiple image registration.

\subsubsection*{Acknowledgements.} The authors gratefully acknowledge TU Delft AI Initiative and Amazon Research Awards for financial support.

% ---- Bibliography ----
% \appendix
% \section{Model Information}
% \input{tables/tab_modelinfo}
%
% ---- Bibliography ----
%
% BibTeX users should specify bibliography style 'splncs04'.
% References will then be sorted and formatted in the correct style.
%\clearpage
\bibliographystyle{splncs04}
\bibliography{bib}

\begin{thebibliography}{10}
\providecommand{\url}[1]{\texttt{#1}}
\providecommand{\urlprefix}{URL }
\providecommand{\doi}[1]{https://doi.org/#1}

\bibitem{arava2021deep}
Arava, D., Masarwy, M., Khawaled, S., Freiman, M.: Deep-learning based motion correction for myocardial t1 mapping. In: 2021 IEEE International Conference on Microwaves, Antennas, Communications and Electronic Systems (COMCAS). pp. 55--59. IEEE (2021)

\bibitem{voxelmorph}
Balakrishnan, G., Zhao, A., Sabuncu, M.R., Guttag, J., Dalca, A.V.: Voxelmorph: a learning framework for deformable medical image registration. IEEE transactions on medical imaging  \textbf{38}(8),  1788--1800 (2019)

\bibitem{bron2013image}
Bron, E.E., van Tiel, J., Smit, H., Poot, D.H., Niessen, W.J., Krestin, G.P., Weinans, H., Oei, E.H., Kotek, G., Klein, S.: Image registration improves human knee cartilage t1 mapping with delayed gadolinium-enhanced mri of cartilage (dgemric). European radiology  \textbf{23},  246--252 (2013)

\bibitem{fu2020deep}
Fu, Y., Lei, Y., Wang, T., Curran, W.J., Liu, T., Yang, X.: Deep learning in medical image registration: a review. Physics in Medicine \& Biology  \textbf{65}(20),  20TR01 (2020)

\bibitem{van2013model}
van~de Giessen, M., Tao, Q., van~der Geest, R.J., Lelieveldt, B.P.: Model-based alignment of look-locker mri sequences for calibrated myocardical scar tissue quantification. In: 2013 IEEE 10th International Symposium on Biomedical Imaging. pp. 1038--1041. IEEE (2013)

\bibitem{haaf2016cardiac}
Haaf, P., Garg, P., Messroghli, D.R., Broadbent, D.A., Greenwood, J.P., Plein, S.: Cardiac t1 mapping and extracellular volume (ecv) in clinical practice: a comprehensive review. Journal of Cardiovascular Magnetic Resonance  \textbf{18}(1), ~89 (2016)

\bibitem{hanania2023pcmc}
Hanania, E., Volovik, I., Barkat, L., Cohen, I., Freiman, M.: Pcmc-t1: Free-breathing myocardial t1 mapping with physically-constrained motion correction. In: International Conference on Medical Image Computing and Computer-Assisted Intervention. pp. 226--235. Springer (2023)

\bibitem{huizinga2016pca}
Huizinga, W., Poot, D.H., Guyader, J.M., Klaassen, R., Coolen, B.F., van Kranenburg, M., Van~Geuns, R., Uitterdijk, A., Polfliet, M., Vandemeulebroucke, J., et~al.: Pca-based groupwise image registration for quantitative mri. Medical image analysis  \textbf{29},  65--78 (2016)

\bibitem{kellman2013t1}
Kellman, P., Arai, A.E., Xue, H.: T1 and extracellular volume mapping in the heart: estimation of error maps and the influence of noise on precision. Journal of Cardiovascular Magnetic Resonance  \textbf{15}(1),  1--12 (2013)

\bibitem{kellman2014t1}
Kellman, P., Hansen, M.S.: T1-mapping in the heart: accuracy and precision. Journal of cardiovascular magnetic resonance  \textbf{16},  1--20 (2014)

\bibitem{klein2009elastix}
Klein, S., Staring, M., Murphy, K., Viergever, M.A., Pluim, J.P.: Elastix: a toolbox for intensity-based medical image registration. IEEE transactions on medical imaging  \textbf{29}(1),  196--205 (2009)

\bibitem{li2021learning}
Li, B., Niessen, W.J., Klein, S., Ikram, M.A., Vernooij, M.W., Bron, E.E.: Learning unbiased group-wise registration (lugr) and joint segmentation: evaluation on longitudinal diffusion mri. In: Medical Imaging 2021: Image Processing. vol. 11596, pp. 136--144. SPIE (2021)

\bibitem{li2023contrast}
Li, X., Zhang, Y., Zhao, Y., van Gemert, J., Tao, Q.: Contrast-agnostic groupwise registration by robust pca for quantitative cardiac mri. In: International Workshop on Statistical Atlases and Computational Models of the Heart. pp. 77--87. Springer (2023)

\bibitem{li2022motion}
Li, Y., Wu, C., Qi, H., Si, D., Ding, H., Chen, H.: Motion correction for native myocardial t1 mapping using self-supervised deep learning registration with contrast separation. NMR in Biomedicine  \textbf{35}(10),  e4775 (2022)

\bibitem{makela2002review}
Makela, T., Clarysse, P., Sipila, O., Pauna, N., Pham, Q.C., Katila, T., Magnin, I.E.: A review of cardiac image registration methods. IEEE Transactions on medical imaging  \textbf{21}(9),  1011--1021 (2002)

\bibitem{martin2020groupwise}
Mart{\'\i}n-Gonz{\'a}lez, E., Sevilla, T., Revilla-Orodea, A., Casaseca-de-la Higuera, P., Alberola-L{\'o}pez, C.: Groupwise non-rigid registration with deep learning: an affordable solution applied to 2d cardiac cine mri reconstruction. Entropy  \textbf{22}(6), ~687 (2020)

\bibitem{messroghli2004modified}
Messroghli, D.R., Radjenovic, A., Kozerke, S., Higgins, D.M., Sivananthan, M.U., Ridgway, J.P.: Modified look-locker inversion recovery (molli) for high-resolution t1 mapping of the heart. Magnetic Resonance in Medicine: An Official Journal of the International Society for Magnetic Resonance in Medicine  \textbf{52}(1),  141--146 (2004)

\bibitem{metz2011nonrigid}
Metz, C.T., Klein, S., Schaap, M., van Walsum, T., Niessen, W.J.: Nonrigid registration of dynamic medical imaging data using nd+ t b-splines and a groupwise optimization approach. Medical image analysis  \textbf{15}(2),  238--249 (2011)

\bibitem{o2022t2}
O'Brien, A.T., Gil, K.E., Varghese, J., Simonetti, O.P., Zareba, K.M.: T2 mapping in myocardial disease: a comprehensive review. Journal of Cardiovascular Magnetic Resonance  \textbf{24}(1), ~33 (2022)

\bibitem{qiao2019fully}
Qiao, M., Wang, Y., Berendsen, F.F., van~der Geest, R.J., Tao, Q.: Fully automated segmentation of the left atrium, pulmonary veins, and left atrial appendage from magnetic resonance angiography by joint-atlas-optimization. Medical physics  \textbf{46}(5),  2074--2084 (2019)

\bibitem{ronneberger2015u}
Ronneberger, O., Fischer, P., Brox, T.: U-net: Convolutional networks for biomedical image segmentation. In: Medical Image Computing and Computer-Assisted Intervention--MICCAI 2015: 18th International Conference, Munich, Germany, October 5-9, 2015, Proceedings, Part III 18. pp. 234--241. Springer (2015)

\bibitem{schelbert2016state}
Schelbert, E.B., Messroghli, D.R.: State of the art: clinical applications of cardiac t1 mapping. Radiology  \textbf{278}(3),  658--676 (2016)

\bibitem{tao2018robust}
Tao, Q., van~der Tol, P., Berendsen, F.F., Paiman, E.H., Lamb, H.J., van~der Geest, R.J.: Robust motion correction for myocardial t1 and extracellular volume mapping by principle component analysis-based groupwise image registration. Journal of Magnetic Resonance Imaging  \textbf{47}(5),  1397--1405 (2018)

\bibitem{tilborghs2019robust}
Tilborghs, S., Dresselaers, T., Claus, P., Claessen, G., Bogaert, J., Maes, F., Suetens, P.: Robust motion correction for cardiac t1 and ecv mapping using a t1 relaxation model approach. Medical Image Analysis  \textbf{52},  212--227 (2019)

\bibitem{wachinger2012simultaneous}
Wachinger, C., Navab, N.: Simultaneous registration of multiple images: similarity metrics and efficient optimization. IEEE transactions on pattern analysis and machine intelligence  \textbf{35}(5),  1221--1233 (2012)

\bibitem{xue2012motion}
Xue, H., Shah, S., Greiser, A., Guetter, C., Littmann, A., Jolly, M.P., Arai, A.E., Zuehlsdorff, S., Guehring, J., Kellman, P.: Motion correction for myocardial t1 mapping using image registration with synthetic image estimation. Magnetic resonance in medicine  \textbf{67}(6),  1644--1655 (2012)

\bibitem{yang2022end}
Yang, J., K{\"u}stner, T., Hu, P., Li{\`o}, P., Qi, H.: End-to-end deep learning of non-rigid groupwise registration and reconstruction of dynamic mri. Frontiers in cardiovascular medicine  \textbf{9},  880186 (2022)

\bibitem{zhang2021groupregnet}
Zhang, Y., Wu, X., Gach, H.M., Li, H., Yang, D.: Groupregnet: a groupwise one-shot deep learning-based 4d image registration method. Physics in Medicine \& Biology  \textbf{66}(4),  045030 (2021)

\bibitem{zhao2023relaxometry}
Zhao, Y., Zhang, Y., Tao, Q.: Relaxometry guided quantitative cardiac magnetic resonance image reconstruction. In: International Workshop on Statistical Atlases and Computational Models of the Heart. pp. 349--358. Springer (2023)

\end{thebibliography}
\end{document}